\documentclass[12pt,preprint]{aastex}

\def\plotfiddle#1#2#3#4#5#6#7{\centering \leavevmode
    \vbox to#2{\rule{0pt}{#2}}
    \includegraphics{#1}}
\newcommand{\Mdot}{\dot M}

\newcommand{\Mstar}{M_*}

\newcommand{\Msun}{M_{\odot}}
\newcommand{\Rin}{R_{\rm in}}
\newcommand{\Rout}{R_{\rm out}}

\newcommand{\Rsun}{R_\odot}
\newcommand{\Msunperyr}{M_{\odot}\,{\rm yr}^{-1}}
\newcommand{\Lsun}{L_{\odot}}
\newcommand{\Lstar}{L_*}

\newcommand{\percc}{\rm \,cm^{-3}}

\newcommand{\persqcm}{\rm \,cm^{-2}}
\newcommand{\gpersqcm}{\rm \,g\,cm^{-2}}

\newcommand{\kms}{\,{\rm km}\,{\rm s}^{-1}}

\def\micron{\hbox{$\,\mu$m}}

\shorttitle{}
\shortauthors{}

\begin{document}


\title{High Resolution K-band Spectroscopy of MWC 480 and V1331 Cyg\altaffilmark{1}}


\author{Joan R. Najita, Greg W. Doppmann}
\affil{National Optical Astronomy Observatory, 950 N. Cherry Ave., 
Tucson, AZ 85719} 

\author{John S. Carr}
\affil{Naval Research Laboratory, Code 7213, Washington, DC 20375}

\and

\author{James R. Graham, J. A. Eisner}
\affil{Astronomy Department, UC Berkeley, Berkeley, CA 94720}


\altaffiltext{1}{The data
presented herein were obtained at the W.M. Keck Observatory, in 
part from 
telescope time allocated to NASA through the agency's scientific 
partnership with the California Institute of Technology and the 
University of California.  The Observatory was made possible 
by the generous financial support of the W.M. Keck Foundation.}


\begin{abstract}
We present high resolution ($R$=25,000--35,000) 
$K$-band spectroscopy of two young stars,  
MWC 480 and V1331 Cyg.  
Earlier spectrally dispersed ($R$=230) 
interferometric observations of MWC 480 
indicated the presence of an excess continuum 
emission interior to the dust sublimation radius, with a spectral shape 
that was interpreted as evidence for hot water emission from 
the inner disk of MWC 480. 
Our spectrum of V1331 Cyg reveals strong emission from CO and 
hot water vapor, likely arising in a circumstellar disk.  
In comparison, our spectrum of 
MWC 480 appears mostly featureless.   
We discuss possible ways in which strong water emission from MWC 480 
might go undetected in our data. 
If strong water emission is in fact absent from the inner disk, as 
our data suggest, the continuum excess interior to the dust 
sublimation radius that is detected in the interferometric data must have 
another origin. 
We discuss possible physical origins for the continuum excess. 
\end{abstract}


\keywords{(stars:) circumstellar matter --- 
(stars:) planetary systems: protoplanetary disks --- 
stars: pre-main sequence ---
(stars: individual) MWC480, V1331 Cyg}



\section{Introduction}

CO overtone emission and hot water emission in the $K$-band 
are both recognized to probe the hot ($>1500$\,K) inner 
gaseous disk surrounding young stars. 
High spectral resolution studies of CO overtone emission have 
been used to probe disks surrounding 
low mass T Tauri stars, intermediate mass Herbig Ae stars, 
and high mass stars 
(e.g., Carr et al.\ 1993; Chandler et al.\ 1993; 
Najita et al.\ 1996; Thi \& Bik 2005; Blum et al.\ 2004; 
Thi et al.\ 2005; Berthoud et al.\ 2007). 
High resolution spectroscopy of water emission, both in the 
$K$-band (Carr et al.\ 2004; Najita et al.\ 2000; Thi \& Bik 2005)
and at longer infrared wavelengths (Salyk et al.\ 2008; Knez et al.\ 2007),  
has also been used to probe the properties of inner gaseous 
disks. 

More recently, it has been argued based on spectrally-dispersed 
interferometric observations that strong water emission is 
responsible for a significant continuum excess inward of the 
dust sublimation radius in a disk surrounding 
a Herbig Ae star (Eisner 2007).  
Several recent interferometric studies have reported similar evidence 
for a bright compact source of near-infrared emission from 
Herbig AeBe stars (Eisner et al.\ 2007; Kraus et al.\ 2007; 
Tannirkulam et al.\ 2008a,b; Isella et al.\ 2008). 
The compact continuum emission is attributed by these 
authors to emission from a gaseous disk interior to the 
dust sublimation radius.  
Such a gaseous disk is expected to show significant 
spectral structure due to emission lines of CO and water and 
other opacity sources (e.g., Muzerolle et al.\ 2004; Eisner 2007). 

Here we report high resolution $K$-band spectroscopy of two 
young stars, MWC 480 and V1331 Cyg. 
MWC 480 (HD 31648; A3pshe+) 
is a Herbig Ae star with a stellar mass of $2.3\Msun$ that is 
surrounded by a rotating disk of gas and dust (Mannings et al.\ 1997).
CO fundamental emission is detected from the source (Blake \& Boogert 2004).
Spatially resolved millimeter and infrared interferometric studies 
of the emission from MWC 480 derive an inclination of 
$i=$26--38 degrees 
for the disk  (Simon et al.\ 2000; Eisner et al.\ 2004). 
Spectrally dispersed interferometric observations find evidence 
for an emission excess located interior to the dust sublimation 
radius with a spectral shape that suggests emission from 
hot water vapor (Eisner 2007). 
Evidence for active accretion in this source includes 
the presence of associated HH objects 
and emission lines of Si IV (1394\AA, 1403\AA) that arise 
from hot accreting gas (Sitko et al.\ 2008; 
also Valenti et al.\ 2000 and Muzerolle et al.\ 2004).  

The other source in our study, V1331 Cyg (LkHa 120), 
is a pre-main-sequence star in the L988 dark 
cloud complex.  
The distance to L988 (see Herbig \& Dahm 2006 for 
a summary) has been estimated 
as 700-800\,pc based on photometry and spectroscopy of nebulous 
sources in the L988 region (Chavarria 1981; 
Chavarria-K \& de Lara 1981). 
Studies of the extinction in the region toward the clouds 
determine a closer distance for the clouds of 550\,pc 
(Shevchenko et al.\ 1991) and $500\pm100$\,pc Alves et al.\ (1998). 

Various studies suggest a spectral type for V1331 Cyg (A8-G5) 
that falls between those of Herbig Ae stars and typical T Tauri stars 
(Kuhi 1964; 
Chavarria 1981;
Hamann \& Personn 1992).  
V1331 Cyg has an associated molecular outflow (Levreault 1988) 
and a massive circumstellar disk 
($0.5\Msun$ assuming 550\,pc distance; McMuldroch et al.\ 1993; 
see also Weintraub et al.\ 1991)  
that is surrounded by a flattened gaseous envelope 
and a ring of reflection nebulosity 
viewed at an inclination of $i \sim 30$ degrees. 
V1331 Cyg is known to show emission in CO overtone bands (Carr 1989). 
High resolution $L$-band spectroscopy that shows water and OH 
emission from V1331 Cyg has also been presented previously 
(Najita et al.\ 2005).

\section{Observations}

A high resolution $K$-band spectrum of V1331 Cyg was obtained by 
one of us (JRG) on 03 July 1999 
as part of the commissioning of NIRSPEC (McLean et al.\
1998), the facility spectrograph on the 10-m Keck 2 telescope.  
The NIRSPEC echelle and cross-disperser gratings were oriented to
image $K$-band orders 33-39 onto the 1024 $\times$ 1024 InSb detector
array through the N7 blocking filter in four 120 second exposures.  Thus
the spectral setup covered 7 spectral regions between $1.93 -
2.31\micron$.  The spectral regions include the $v$=2--0 CO overtone lines
at the long wavelength end, the Br $\gamma$ line in the middle of
the $K$-band, and numerous lines of water throughout.  

Using the
$0\farcs 288 $ (2-pixel) wide slit, $R \simeq$ 35,000 ($8.5 \kms$)
spectra were acquired in four beam positions along the $12 \arcsec$
long slit in $\sim$1$\arcsec$ seeing (FWHM $K$-band).  During
these observations, the internal instrument rotator was on,
keeping the slit at a fixed position angle on the sky while the
non-equatorially mounted telescope tracked.
To remove telluric absorption features, spectra
of an early-type star (HR 8146; B2V), 
were obtained with the same grating setting and 
at a similar air
mass following the V1331 Cyg observations.  Spectra of the internal
NIRSPEC continuum lamp were taken for flatfielding, and spectra of
internal arc lamps (with emission lines of Argon, Krypton, Xenon,
and Neon) were obtained for wavelength calibration.

Similar $K$-band spectra of MWC 480 were obtained on 03 January 2007 
with NIRSPEC.  The spectral coverage overlapped that of the 
V1331 Cyg observations by $\sim$75\% in orders 33--38.  
The MWC 480 spectra were acquired through the $0\farcs
432 $  (3-pixel)  wide slit ($R \simeq$ 24,000; $12.5 \kms$ resolution) 
in $K$-band seeing of  $1\farcs 3$ (FWHM).  Eight exposures of 40s
each were taken in an ABBA nod pattern in which the object was placed 
at $\pm 6\arcsec$ from the center
of the 24$\arcsec$ long slit. 
The observations were made in stationary mode. 
These 
observations made use of the newer NIRSPEC detector array which
has improved performance and fewer bad pixels than the 
detector used in the 1999 observations.
There was some wind shake in the east-west direction 
(i.e., across the slit) during our observations. 

\section{Data Reduction}

\subsection{V1331 Cyg}

As a first pass, bad pixels present in the raw object, telluric, and
calibration source frames  were identified and removed by interpolation
using the ``fixpix" algorithm within the REDSPEC reduction package
(a custom echelle 
reduction package developed by Kim, Prato, and McLean).
Standard IRAF packages (Massey et al.\ 1992; Massey
1997) were then used to reduce the data.  Differencing pairs of exposures
taken in different beam positions (e.g., beam A - B, and beam C - D) 
was adequate for sky subtraction.  The differenced pairs were
then flatfielded by dividing by the internal continuum lamp frame.  
Object and telluric calibration spectra in
each echelle order were extracted from a spatial profile that was
fixed at $\pm$2.5 pixels ($\pm 0\farcs5$) about the profile peak.
The object and telluric spectra extracted at each nod position were 
wavelength calibrated using selected telluric absorption lines 
(for orders 33, 34, and 35) 
with wavelengths taken from the HITRAN database
(Rothman et al.\ 1998), and emission lines of Argon, Krypton, Xenon,
and Neon from exposures in the internal lamp spectra 
(for orders 36, 37, and 38).

Telluric features in each of the four nod positions were removed
from our object spectra by dividing by the spectrum of the telluric
standard star obtained at a similar nod position.  
Order 35 contained emission from Br$\gamma$ 
in our telluric standard.  
To avoid introducing spurious structure in the object spectra, 
we divided by the telluric standard to correct for telluric 
absorption only in regions far from the Br$\gamma$ line.   
Residual
telluric features are therefore present in the spectral region within
$\sim$415~$\kms$ of the Br $\gamma$ line center 
(2.1638 -- 2.1693$\micron$).

Wavelength calibrated
spectra at different nod positions in each order were summed and
then multiplied by a blackbody of 22,000\,K to restore the true
continuum shape after division by the B2V telluric standard.
The flux level in the V1331 spectrum was estimated by the conversion
from observed counts in the observations of the standard to its
2MASS $K$-band magnitude ($m_K$ = 4.48; 10\,Jy), and assuming an 
equal slit loss between the standard and object observations.

\subsection{MWC 480}

The MWC480 observations were reduced using a similar procedure. 
Since these observations made use of the
improved NIRSPEC array, there were significantly fewer bad pixels.  
Thus, bad pixels were individually identified and their
values fixed by interpolation using the IRAF task ``fixpix".
The observations of MWC 480 were taken with a wider slit (3-pixel,
0$\farcs$432) and in windy conditions with poor seeing, compared
to V1331 Cyg.  As a result, spectral images of MWC 480 and the
telluric standard star (HR 1412) that were observed at similar beam
positions were summed together without offsets, and their spectra
were extracted from a relatively wide ($\pm$5 pixels) spectral
profile.
Good signal-to-noise in the exposures of the arc lamps (i.e. Ar,
Kr, Xe, and Ne) permitted wavelength calibration of all orders
except order 33, where telluric absorption lines were used.

Telluric features in the MWC 480 spectra were removed by dividing 
by the spectrum of the standard star (HR 1412) obtained at a 
similar nod position.  Several weak stellar absorption lines were 
present in the standard star spectrum.  These were modeled 
and removed using a stellar synthesis model with 
$T_{\rm eff}=7600$\,K and $\log g = 3.5$, to match the 
A7 III spectral type of HR 1412
(Doppmann, Najita, and Carr 2008).  Since order 35 contained broad
Br $\gamma$ absorption in our standard, 
we only divided by the telluric standard to correct for telluric 
absorption in regions 
$\gtrsim 450 \kms$ from the Br $\gamma$ peak. 
Thus, telluric features are
present in the object spectrum between 2.163 - 2.170 $\mu$m.

To restore the true continuum slope in MWC 480, the spectra were
multiplied by a 7600\,K blackbody after division by the telluric
standard.  The flux level of MWC 480 was calibrated using the
observed counts and 2MASS $K$-band magnitude of HR 1412 
($m_K = 2.880$ or 47\,Jy), assuming
equal slit losses in the object and telluric observations.

\section{Results}

Figures 1a-1f compare the spectra of V1331 Cyg and MWC 480. 
The spectrum of V1331 Cyg shows significant structure 
in all orders.  
In order 33 (Fig.\,1a), strong 
CO overtone ($v$=2--0) bandhead emission is present as are 
numerous weaker emission lines blueward of the bandhead. 
Numerous narrow emission features are present in all other 
orders, as is strong Br $\gamma$ emission.  
The wavelengths of many of the narrow emission features agree with 
the expected wavelengths of lines identified as water in the 
infrared sunspot spectrum (Polyansky et al.\ 1997). 
In contrast, the MWC 480 spectrum appears mostly featureless 
in all orders, with the exception of strong Br $\gamma$ 
emission in order 35 (Fig.\,1c; see also Appendix).

\subsection{Lack of Stellar Photospheric Features from MWC 480}

The lack of stellar photospheric features in the MWC 480 
spectrum is consistent with the expected strength of such 
features given the stellar spectral type and photometric 
veiling estimated for the source. 
Figure 2 shows the spectral energy distribution for MWC 480
determined from optical photometry from the Tycho-2 catalog 
(H{\o}g et al.\,2000), the USNO B catalog (Monet et al.\, 2003), 
and infrared photometry from 2MASS (Skrutskie et al.\ 2006) 
and Morel \& Magnenat (1978), 
dereddened ($A_V = 0.3$, Mannings and Sargent 1997) using an
interstellar extinction law with $R_V=3.1$ (Mathis 1990).
A 8400\,K stellar photosphere, consistent with a stellar spectral type 
of A4 (Simon et al.\ 2000), provides a good fit to 
the dereddened optical photometry, assuming a stellar luminosity
$\Lstar = 22 \Lsun$ and a distance of 170\,pc.  
A distance of 170\,pc is within 2-$\sigma$ of the Hipparcos value 
(van den Ancker et al.\ 1998). 
At a distance of 170\,pc, 
dynamical constraints imply a stellar mass of $2\Msun$ for MWC 480 
(Simon et al.\ 2000), consistent with the stellar mass inferred 
from the pre-main-sequence tracks of Siess (2000) for 
this luminosity and spectral type.

The 2MASS photometry of MWC 480 shows an infrared SED rising well
above the stellar photosphere; the $K$-band excess 
is $\simeq 4$ times the stellar photospheric continuum.  
A continuum veiling of this magnitude is consistent with the 
absence of stellar photospheric features in the NIRSPEC spectra. 
To estimate the stellar contribution to the spectrum, we 
generated stellar synthetic spectra 
using the program MOOG (Sneden 1973) and Hauschildt et al.\ (1999)
model atmospheres.  The initial line list was
taken from Kurucz (1993), and individual line parameters were
adjusted to fit the observed disk-center solar spectrum of
Livingston \& Wallace (1991). 

A synthetic stellar photosphere with 
$T_{\rm eff} = 8400$\,K, $\log g = 4.5,$ and $v \sin i = 75\kms$ 
predicts weak ($< 2.7$\% deep) stellar absorption lines in the 
$K$-band (e.g. Mg I at $2.28143\micron$).
A stellar rotation velocity of $v\sin i = 75\kms$ is typical, 
or even conservative, for a $\sim 1$\,Myr old A-type star,
based on rotation statistics found for the Orion OB association 
(Wolff, Strom, \& Hillenbrand 2004).   
With a continuum veiling of $r_K = 4$, 
the ratio of the continuum excess to the stellar photospheric 
flux at $K$,
these features would be hidden within the noise of our 
observations ($S/N \sim 400$), consistent with the absence of 
absorption features in our spectra of MWC 480.

\subsection {CO and Water Emission from V1331 Cyg} 

In the V1331 Cyg spectrum, 
the two strongest features blueward of the $v$=2--0 CO bandhead have 
a FWHM of $18-22\kms$. 
These lines are blended with weaker water lines, although 
each feature is dominated by a single strong water line. 
Isolated CO lines redward of the bandhead have 
FWHM $\approx 29-35\kms$. 
Thus, the water emission lines are narrower than the CO lines, 
similar to the situation found for other sources such as 
SVS-13 and DG Tau (Carr et al. 2004; Najita et al.\ 2000).  

CO and water emission from the young star SVS-13 
was previously studied by Carr et al.\ (2004) 
in the region near the $v$=2--0 CO bandhead at $2.3\micron$. 
As discussed in Carr et al.\ (2004), a major difficulty in 
modeling near-infrared water emission is the need for a complete 
and accurate line list.  In their study, Carr et al.\ (2004) 
developed a water line list for the $2.3\micron$ region. 
They started with the Partridge \& Schwenke (1997) theoretical 
line list and improved the accuracy of both the line positions 
and strengths. 
Using the derived line list, they fit the spectrum of SVS-13 
with a simple model of a differentially rotating disk. 

Here, we adopt the water linelist and disk model used by 
Carr et al.\ (2004) to model the spectrum of V1331 Cyg. 
As in Carr et al.\ (2004), the observed spectrum is 
modeled 
under the assumption of Keplerian rotation and thermal 
level populations (see also Carr et al.\ 1993; Najita et al.\ 1996).  
The disk model parameters include a stellar mass, 
the projected rotational velocity $v\sin i$ at the inner radius 
of the emission, 
the inner and outer radii of the emission, 
and the temperature and 
column density distribution 
for the emitting gas, both parameterized as power laws in radius.
The distance and observed velocity shift are additional 
parameters.  

Since the stellar properties of V1331 Cyg (distance, spectral
type, extinction) are poorly known, the modeling that we present
for the molecular emission from V1331 Cyg is not meant to provide
a definitive interpretation of the emission from the source.  Instead,
the modeling has the limited goal of demonstrating that the observed
emission features are consistent with water and CO emission from a
differentially rotating disk.  A more detailed discussion of the
properties of the molecular emission will be presented in a future
publication.

In this context, 
we assume a stellar mass of $1.8\Msun$ which is consistent 
with estimates for the bolometric luminosity of the source 
($35-55\Lsun$; Hamann \& Persson 1992; Shevchenko et al.\ 1991) 
and the range of spectral types estimated in the literature 
(A8--K2; Chavarria 1981; Hamann \& Persson 1992). 
We also assume a distance of 550\,pc (Shevchenko et al.\ 1991; 
Alves et al.\ 1998). 
With the combination of stellar spectral type and extinction 
preferred by Chavarria (1981; F0 and $A_V=2.4$) 
and our assumed stellar luminosity of $12 \Lsun$, the stellar 
radius is $2.2\Rsun$.  

Although the CO and water emission may arise in a vertical 
temperature inversion region in the disk, for consistency with our 
limited goal for the modeling in this paper, we ignore the 
continuum in the modeling. 
The spectral line emission is therefore assumed to arise in a 
vertically-uniform, continuum-free layer.  
We also do not include an underlying continuum in fitting the 
line emission.  

The ratios of the water lines blueward of the bandhead indicates 
that the water emission is optically thick.
Very optically thick emission (a line center optical depth of 
$\tau \sim 15$ for the water line at $2.2916\micron$) 
is needed to reproduce the strength of the strong water lines 
(e.g., at $2.2916 \micron$ and $2.2928\micron$) 
relative to the weaker lines 
(e.g., at $2.2907\micron$ and $2.2912\micron$).
Figure 3 shows a fit to the water lines alone using a model 
of emission from a rotating disk.  
The model parameters for this fit are 
a temperature of 1500\,K 
and a line-of-sight column density of $6.8\gpersqcm$. 
The assumed parameters correspond to a constant line-of-sight 
water column density of $1.4\times 10^{21}\persqcm$ 
over the emitting region.

The emission lines are narrow, requiring a low projected disk 
rotational velocity $v\sin i$. 
Fitting the width of the emission requires a rotational velocity 
at an inner radius $\Rin$ of $v\sin i =14\kms$ 
when smoothed to the $8.5\kms$
resolution of the 2-pixel NIRSPEC slit 
and assuming no microturbulent broadening. 
The strength of the emission can be fit assuming 
an annular emitting region extending from an 
inner radius of $\Rin = 5.5\Rsun$ to an outer radius $\Rout = 5\Rin$. 
For the assumed stellar mass and the low value of $v\sin i$ 
required at the inner radius, $\Rin = 5.5\Rsun$ corresponds 
to a face-on inclination of 3 degrees. 
The observed emission features are well fit with a topocentric 
velocity of $-29\kms$, in good agreement 
with the system radial velocity 
($v_{\rm LSR} = -0.7\kms$; Levreault 1988; McMuldroch et al.\ 1993). 

We can obtain a reasonable fit to both the CO and water emission  
using the differentially rotating disk model (Figure 4). 
The model parameters for this fit are 
a radial temperature distribution  
$T = 4100 (r/\Rin)^{-0.55}$ and 
a line-of-sight column density that varies as 
$\Sigma = 40 \gpersqcm (r/\Rin)^{-0.6}$ 
between an inner radius of $\Rin = 2.3\Rsun$ and 
an outer radius of $\Rout = 15\Rin$. 
The line-of-sight rotational velocity at $\Rin$ is   
$v\sin i =23\kms,$ and $4\kms$ of 
Gaussian microturbulent broadening is assumed throughout.
The $v\sin i$ and the microturbulence are jointly constrained by 
the widths of the isolated water and CO lines, 
the beating of the CO lines immediately redward of the bandhead, 
and 
the strengths of the CO lines far from the bandhead compared to 
the strength of the bandhead. 
The $4\kms$ local line broadening used in the fit is in 
excess of the thermal dispersion of $0.9\kms$ for CO at 
2500\,K and comparable to the sound speed at that temperature. 

Since the relative strength of the CO emission compared to the water 
emission is much larger than expected for a CO-to-H$_2$O 
abundance ratio in chemical equilibrium, 
the water abundance in the fit is therefore scaled down by a 
factor of 0.2 relative to its chemical equilibrium value. 
Because the water emission is less optically thick in this model 
than in the model shown in Figure 3, the weaker water lines 
(e.g., at $2.2907 \micron$ and $2.2921 \micron$)
are underfit in the model compared to Figure 3.

In the model, the CO emission forms over radii $\sim 1-7\Rin$, 
whereas the water emission forms at larger radii $\sim 2.2-7.5\Rin$. 
Since the CO emission extends in to smaller disk radii (and higher
disk rotational velocities) than the water emission, the water lines
are narrower than the CO lines in the synthetic spectrum, in agreement
with the observations.

As a caveat, we note that 
for a source with the properties assumed above,
the SED of V1331 Cyg would imply significant veiling, 
with an optical excess that is 1.7 times the strength of the 
stellar continuum. 
Significant veiling at optical wavelengths may help to explain 
the lack of stellar absorption lines in the optical spectrum of 
this source (Chavarria 1981). 
If V1331 Cyg actually experiences little or no optical 
veiling, the SED would be better fit with a stellar luminosity 
of $28\Lsun$ and a stellar radius of $3.4\Rsun$.  
In such a case $\Rin$ would have to be larger, and the other disk 
model parameters could be modified to 
produce fits comparable to those shown in Figures 3 and 4.
We could fit the water emission alone (the equivalent of Figure 3) 
by adopting a somewhat smaller disk column density of 
$4\gpersqcm$ for a larger inner radius of $6.5\Rsun$.   
We could fit both the CO and water emission (the equivalent of 
Figure 4) by adopting a cooler temperature distribution 
$T=3400(r/\Rin)^{-0.55}$ for a larger inner radius of $4\Rsun$.   

The column density adopted in the modeling is much less than 
that of either
the minimum mass solar nebula or a steady accretion disk that
accretes at a rate typical of T Tauri stars.  The smaller column
density adopted in the modeling may result because the molecular
emission arises in a temperature inversion region in the disk
atmosphere rather than from the entire vertical column density of
an optically thick disk.  Alternatively, or in addition, CO and
water may not be present throughout (or sufficiently warm enough
to emit over) the entire vertical column density of the disk.

The possibility of non-thermal line broadening in the disk 
atmosphere, as indicated by the shape of the CO overtone emission, 
is similar to the result obtained by Carr et al.\ (2004) for 
the young star SVS-13.  In the earlier study, the magnitude 
of the non-thermal local line broadening was $\sim 11 \kms$ 
for a Gaussian profile, larger than both the $4\kms$ used here 
for V1331 Cyg and the line broadening found for the 
CO bandhead emission from the Herbig Ae stars WL16 and 
1548c27 (Najita et al.\ 1996).  
We have previously speculated that non-thermal 
line broadening is a consequence of turbulence in the 
disk atmosphere, perhaps produced by the magnetorotational 
instability. 

The strong CO emission compared to water emission that we 
find for V1331 Cyg has also been found for 
SVS-13 (Carr et al.\ 2004) and the $6\Msun$ source 
studied by Thi \& Bik (2005). 
These results have been interpreted as evidence for 
a reduced H$_2$O/CO abundance ratio in the disk atmosphere. 
A reduced H$_2$O/CO ratio is indeed found for models 
of strongly irradiated disks surrounding high mass stars 
(Thi \& Bik 2005).  
An alternative interpretation 
might consider the possibility that the water emission 
is reduced in strength because it arises from a 
significantly narrower emitting area than we have assumed.  
This modeling approach has been used in fitting the 
(spectrally unresolved) rich mid-infrared molecular emission 
of classical T Tauri stars (Carr \& Najita 2008). 
The near face-on geometry of the V1331 Cyg disk allows for 
this possibility despite the high velocity resolution of our  
observations. 
Detailed theoretical studies of the thermal-chemical 
properties of disk atmospheres would be needed to determine 
whether a restricted annular emission region is a reasonable 
interpretation.

\subsection{Lack of Water Emission from MWC 480?}

Compared to the rich emission line spectrum of V1331 Cyg, 
the MWC 480 spectrum appears mostly featureless.  
The lack of obvious water emission features is perhaps surprising 
given the results of Eisner (2007).  
We can use the V1331 Cyg spectrum as a ``water emission template'' 
to determine whether 
velocity broadening by disk rotation or other line broadening 
processes can smear out the spectral signature of water in 
the MWC 480 spectrum. 

In Eisner (2007), the observed spectrum and spectrally dispersed 
visibilities were best fit with a model that included emission from 
(1) the star, 
(2) a ring of optically thick dust with a radius 0.28\,AU and a 
temperature of 1200\,K,
(3) a hot compact component that was modeled as blackbody emission 
from a ring with a radius $<0.1$\,AU and a temperature of 2410\,K, and 
(4) a ring of gaseous water with a radius of 0.16\,AU, 
a temperature of 2300\,K, and a column density of $1.2\times 10^{19}\persqcm$.
At $2.1\micron$, these four components contribute 
approximately 1\,Jy, 1.7\,Jy, 1\,Jy, and 0.1\,Jy, respectively. 
We can consider two cases: 
water emission might account for 
both components (3) and (4) or for component (4) only. 

The rotational broadening of the water emission is 
expected to be $50-70\kms$ based on 
the stellar mass of MWC\,480 ($2.3\Msun$), 
the measured inclination $i=$26--38 degrees of the MWC 480 disk  
(Simon et al.\ 2000; Eisner et al.\ 2004) 
and the disk radius (0.16\,AU) of the water 
emission in the model of Eisner (2007). 
For the hot compact component, 
which had a radius $\lesssim 0.1$\,AU, the expected 
rotational broadening is somewhat larger, 
$\gtrsim 60-90\kms$. 
Similar velocity widths are found for the CO 
fundamental emission from MWC 480, which is also believed 
to originate in the disk.  
The emission has a FWHM of $62 \kms$ for the $J< 9$ lines 
and $\sim 80\kms$ for the $J>25$ lines 
(Blake \& Boogert 2004). 

To explore the possibility of rotational broadening, 
we looked at 
order 36, a spectral order that experiences limited 
telluric absorption and which is located toward the short 
wavelength end of the $K$-band where the water emission is 
expected to become significant. 
Figure 5 (top panel) shows the (continuum subtracted) 
emission line structure in this order of the V1331 Cyg spectrum, 
scaled in flux so that when it is smoothed to a 
resolution of $R=230$ ($1300 \kms$; horizontal line)  
the equivalent pseudo-continuum excess is $\sim 1$\,Jy, 
the flux of the hot compact component 
at $2.1\micron$.  
We also show the line emission spectrum (boxcar) 
smoothed by 
36 pixels (or $\sim 160\kms$; red line), 
larger than the expected 
rotational broadening for either the water or the hot 
compact component. 

A continuum level of 2.7\,Jy is added to the smoothed 
line emission spectra to bring the total average 
flux level up to the continuum level of 3.7\,Jy 
found for MWC 480 by Eisner (2007). 
The middle panel of Figure 5 shows the resulting 
spectrum (red line) overplotted on the observed MWC480 
spectrum in order 36 (black line).  
Significantly more structure is present than in the 
MWC 480 spectrum.  Thus, rotationally broadened water emission 
seems unlikely to account for the flux attributed to the 
hot compact component in the Eisner (2007) model. 

The bottom panel of Figure 5 shows an equivalent plot for 
the case where 
the pseudo-continuum contributed by the water emission 
is smaller, 0.1\,Jy, the strength of the flux attributed 
to component (4) 
in the Eisner (2007) model.  In this case as well, the 
MWC 480 spectrum shows less structure than in the model. 
Thus, water emission seems unlikely be present at the level 
needed to account for the flux attributed to either the 
hot compact component or the water component of the 
Eisner (2007) model. 

A possible caveat is that the water emission reported by 
Eisner (2007) may be time variable.  
The $K$-band flux of MWC 480 is known to vary with time 
by a modest amount (15\%; Sitko et al.\ 2008; de Winter et al.\ 2001). 
However, the strength of the $K$-band continuum in our spectrum is
similar to that reported by Eisner (2007), which does not suggest
a large spectral variation between the two epochs of observation.

Another possible caveat is that V1331 Cyg is not a good 
template for the water emission in MWC 480. 
The water emission in MWC 480 may be much more highly 
optically thick and line blanketed, involving much 
higher water column densities than those considered 
by Eisner (2007) or than is present in the V1331 Cyg spectrum.  
The model of water emission used by Eisner (2007) 
to fit the interferometer observations of MWC\,480 is 
far from being line blanketed.  It 
is in fact more optically thin than the water emission 
observed in the V1331 Cyg spectrum.  The model parameters 
used by Eisner (2007), 
a temperature of 2300\,K and a column density of 
$1.2\times 10^{19}\persqcm$, 
do a reasonable job reproducing the relative fluxes of many of 
the water lines in the region immediately blueward of the 
CO overtone bandhead, 
with the important exception that the lines at 
2.2917 and $2.2928\micron$ are factors of 2--3 too 
strong compared to the weaker water lines.  This is because 
the water emission in the Eisner (2007) model is optically 
thin, whereas the relative fluxes of the water lines in the 
V1331 Cyg spectrum indicate more optically thick emission.

A more detailed study of possible water emission in MWC 480 
would benefit from the development of a reliable water line list 
for the short wavelength end of the $K$-band.  This is a subject 
for future work.

\section{Discussion}

V1331 Cyg shows a rich spectrum of water emission in 
the $K$-band.  
In comparison, the $K$-band spectrum of MWC 480 shows little 
emission from water or any other spectral lines.  
The non-detection of water emission is consistent with the 
absence of CO overtone emission from MWC 480: 
all sources of $K$-band water emission 
reported in the literature also show CO overtone emission 
(Carr et al.\ 2004; Thi \& Bik 2005; Najita et al.\ 2000; 
this paper; see also van Boekel 2007).  

The spectral shape of the low-resolution Keck interferometric 
spectrum of MWC 480 (Eisner 2007) is consistent with 
the lack of water emission features in the high-resolution 
spectrum of MWC 480.
One difference between the low-resolution spectrum 
of MWC 480 and the spectrum of a source like SVS-13, which is 
known to show strong water emission,  
is that at low spectral resolution SVS-13 shows an upturn in 
the continuum at both the long- and short-wavelength ends of 
the $K$-band.  The spectral shape is interpreted as a signature of 
water emission from the $1.9\micron$ and $2.7\micron$ 
water bands (Carr et al.\ 2004).  
In contrast, the spectrum of MWC 480 shows a significant excess
over the continuum at the short wavelength end, while the long-wavelength
spectrum is consistent with continuum emission only. 
To fit the data, Eisner (2007) used hot (2300\,K) water emission, 
which produces relatively more flux at the short-wavelength end 
of the $K$-band than would cooler material.

Although MWC 480 shows neither CO overtone nor water emission, 
there is evidence for disk gas 
inward of the dust sublimation radius, 
as would be expected for a source with a significant outer 
disk ($0.02-0.2\Msun$; Mannings \& Sargent 1997; 
Hamidouche, Looney, \& Mundy 2006) that is believed to be  
accreting (Sitko et al.\ 2008). 
The high-$J$ CO fundamental emission from MWC 480 has line wings 
extending to $\sim 80 \kms$ (Blake \& Boogert 2004).  
At an inclination of $i=26-38$  and 
a stellar mass of $2.3\Msun$, this implies that the CO 
fundamental emission extends to within $0.06-0.12$\,AU 
for Keplerian rotation, well within  
the dust sublimation radius of $\sim 0.28$\,AU 
(Eisner 2007).  
While there is clearly gas within the dust sublimation radius 
of MWC 480, the CO and water column densities and/or the 
temperature there are perhaps too low to produce 
CO overtone or water emission. 

The presence of CO fundamental emission, accompanied by the 
absence of CO overtone and $K$-band water emission, is the 
situation commonly found for classical T Tauri stars.  
CO fundamental emission is detected from nearly all accreting 
T Tauri stars studied (Najita et al.\ 2003), whereas CO overtone 
and $K$-band water emission are restricted to the high accretion 
rate systems (Najita et al.\ 2000, 2007). 
CO overtone emission is also rare among high mass stars 
(Hanson et al.\ 1997). 

In contrast, models of gaseous inner disks  
that assume LTE molecular abundances and include 
NIR opacity sources such as water, CO, and H$^-$ 
predict prominent near-infrared spectral structure 
from Herbig Ae disks. 
In Muzerolle et al.\ (2004), a dust-free 
gaseous inner disk is found to be capable of producing a 
significant $K$-band flux.  At a nominal Taurus distance of 140\,pc, 
an excess of $\sim 1$\,Jy in the $K$-band, the flux of 
the hot compact continuum in MWC 480, can be produced in disks 
with accretion rates of $\sim 10^{-8}\Msunperyr$.  
Although such a model can plausibly account for the magnitude 
of the compact excess that is observed in MWC 480, the excess 
is also predicted to show strong $K$-band emission features of 
CO and water, which are not observed.  

The models further predict prominent CO and water features 
over a wide range in accretion rate 
($10^{-8}-10^{-5}\Msunperyr$; 
Muzerolle et al.\ 2004; Calvet et al.\ 1991).
Emission is predicted at lower accretion rates 
and absorption at higher accretion rates. 
The relative rarity of CO overtone and water features 
(in emission or absorption)
from Herbig Ae stars suggests that something is missing 
from the assumed physical-thermal-chemical structure of 
the disk in these models.  Possibilities include non-LTE 
abundances, photodissociation,   
and missing sources of continuum opacity and ionization.

If water emission does not appear to be responsible for the 
excess detected inward of the dust sublimation radius in 
MWC 480, what is the origin of the excess?   
The lack of molecular emission might be explained 
by the presence of a competing source of continuum opacity 
that produces the hot compact excess. 
Possible additional opacity sources are 
high temperature condensates, H$^-$, and free-free emission.  
Materials such as corundum (Al$_2$O$_3$), 
hibonite (CaAl$_{12}$O$_{19}$), 
perovskite (CaTiO$_3$), and gehlenite (Ca$_2$Al$_2$SiO$_7$) 
are found in CI chondrites and 
have high sublimation temperatures of 1640--1800 K 
(Posch et al.\ 2007).  
However, they have poor $K$-band emission efficiencies 
$Q_{\rm abs} \lesssim 10^{-3}$ (Henning et al.\ 
1991\footnote{http://www.astro.uni-jena.de/Laboratory/OCDB/oxsul.html}; 
also B. Sargent, personal communication), 
which makes it less 
likely that they will contribute significantly to the $K$-band opacity. 

Graphite grains, often invoked to explain the 2175\AA\ bump in
the interstellar extinction curve, also have a high sublimation
temperature ($> 2000$\,K) at interstellar pressures (Krugel
2003; Salpeter 1977).  However, graphite may be destroyed by processes
such as chemisputtering at much lower temperatures $\sim 1200$\,K
(Lenzuni, Gail, \& Henning 1995; Duschl, Gail, \& Tscharnuter 1996).
So it is unclear whether it can explain the observed hot, compact
excess in MWC 480.
Another possibility is that solids disappear through sublimation
over a finite range in temperature (and disk radius).
For example, in considering the balance between the solid and
gas phases for magnesium-iron silicates, Duschl et al.\ (1996)
found that the gas and dust phases coexist over a range in
temperature, with a fraction of silicates surviving to temperatures
$\sim 2100$\,K at the densities of the inner disk region ($\sim 0.1$\,AU).
Aluminum-calcium-silicates may be sustained to somewhat
higher temperatures.
Detailed modeling of this kind may be needed to understand
the conditions under which dust grains may contribute a modest
residual continuum opacity in the high temperature inner regions of disks.

A significant free-free continuum flux requires a significant 
electron abundance in the inner disk. 
For optically thin free-free emission from a 
$10^4$\,K gas in a cylindrical volume that is 
$r =0.1$\,AU in radius and $h = 0.1r$ in height, 
an electron density of $n_e \sim 3\times 10^{11}\percc$ 
is needed to produce a continuum flux of $\sim 1$\,Jy 
at a distance of 140\ pc.
In comparison, 
for a steady accretion disk accreting at the rate $\Mdot$, 
the disk column density is $\Sigma = \Mdot/3\pi\alpha c_s H$
where $\alpha$ is the viscosity parameter, and 
$c_s$ is the sound speed. 
The quantity $H = c_s/\Omega$ is the disk scale height, 
where $\Omega$ is the Keplerian angular velocity. 
The average disk density can be estimated as 
$\bar n_{\rm H} = \Sigma/\mu m_H H \simeq 5\times 10^{15}\percc$
for an accretion rate of $10^{-7}\Msunperyr,$ 
a stellar mass $\Mstar = 2.3\Msun$, 
and a viscosity parameter $\alpha = 0.01$. 
Comparing these densities, we require a high average electron fraction 
in the disk $\bar x_e = n_e / \bar n_{\rm H} \sim 10^{-4}$ 
in order to produce a $\sim 1$\,Jy free-free continuum 
flux from a disk accreting at $10^{-7}\Msunperyr$.

Continuum opacity from H$^-$ is another possibility. 
At a column density $N_{\rm H} > 10^{26}\persqcm$ and 
temperature $T > 4000$\,K, 
the H$^-$ continuum optical depth $\tau > 1$ 
in chemical equilibrium (e.g., Najita et al.\ 1996). 
Such warm temperatures are expected for inner disks 
with accretion rates $>10^{-7}\Msunperyr$; 
the midplane H$^-$ opacity is expected to become 
significant under these conditions (Muzerolle et al.\ 2004). 
The possibility of a high H$^-$ opacity at  
a column density $N_{\rm H} > 10^{26}\persqcm$ is also 
noted by Thi \& Bik (2005). 
Further studies of the gas temperature, electron fraction, 
and H$^-$ abundance expected for the specific conditions in the 
inner region of the MWC 480 disk would be useful to 
sort out these possibilities.

\section{Summary and Future Directions}

Several recent interferometric studies have reported evidence 
for a bright compact source of near-infrared emission interior 
to the dust sublimation radius in Herbig AeBe stars 
(Eisner 2007; Eisner et al.\ 2007; Kraus et al.\ 2007; 
Tannirkulam et al.\ 2008a,b; Isella et al.\ 2008; Akeson et al.\ 2005). 
The compact continuum emission has been attributed 
to emission from a gaseous disk and variously modeled as 
hot water emission (Eisner 2007), 
a uniform disk of a given flux (Tannirkulam et al.\ 2008a), 
a 2500\,K blackbody (Isella et al.\ 2008), or 
an optically thick disk that reradiates away the accretion energy 
(Kraus et al.\ 2007).  
Such a gaseous disk is expected to show significant 
spectral structure (e.g., Muzerolle et al.\ 2004; Eisner 2007). 

Our study of MWC 480 does not reveal any significant spectral 
structure in the $K$-band spectrum of the source, suggesting 
that the compact excess is produced by a continuum process 
(e.g., high temperature condensates, H$^-$, or free-free emission) 
rather than by line emission.  
It would be interesting to carry out similar studies for 
other sources with compact continuum excesses in order to 
determine if the results obtained here for MWC 480 apply 
more broadly to other sources with compact continuum 
excesses. 

Given its narrow emission lines, V1331 Cyg is a good template 
for warm water emission from a circumstellar disk.  The emission 
spectrum may also be useful in constructing a reliable water 
line list for the $K$-band.






\acknowledgments

We thank Ben Sargent and Michael Meyer for helpful insights on 
high temperature condensates. 
Financial support for this work was provided by the NASA Origins 
of Solar Systems program (NNH07AG51I) and 
the NASA Astrobiology Institute under Cooperative Agreement 
No.\ CAN-02-OSS-02 issued through the Office of Space Science.
This work was also supported by the Life and Planets Astrobiology 
Center (LAPLACE). 
Basic research in infrared astronomy at the Naval Research Laboratory
is supported by 6.1 base funding. 
This publication makes use of data products from the Two Micron All
Sky Survey, which is a joint project of the University of Massachusetts
and the Infrared Processing and Analysis Center/California Institute of
Technology, funded by the National Aeronautics and Space Administration
and the National Science Foundation.
The authors wish to recognize and acknowledge the very significant
cultural role and reverence that the summit of Mauna Kea has always
had within the indigenous Hawaiian community.  We are most fortunate
to have the opportunity to conduct observations from this mountain.

\appendix
\section{Br$\gamma$ Emission}

As noted in \S 3, the spectra of V1331 Cyg and MWC 480 
were not corrected for telluric absorption 
in the region immediately surrounding the Br$\gamma$ line 
because of the presence of Br$\gamma$ absorption in the 
telluric standard.  We can nevertheless deduce a few basic 
features of the Br$\gamma$ emission present in both spectra. 
The emission is broad in both V1331 Cyg and MWC 480, with 
FWZI widths of $550\kms$ and $640\kms$ respectively.  
The emission equivalent width and integrated line flux 
are approximately $-9.8$\AA\ and 
$1.8\times 10^{-16}{\rm \,W\,m^{-2}}$, respectively, 
for V1331 Cyg and approximately $-6.9$\AA\ and 
$1.5\times 10^{-15}{\rm \,W\,m^{-2}}$, respectively, 
for MWC 480.

The emission centroids are blueshifted from their 
systemic velocities by $-12.7\kms$ for V1331 Cyg and 
$-6.8\kms$ for MWC 480.  Similarly blue-shifted Br$\gamma$ 
centroids are found for lower mass T Tauri stars (e.g., Najita 
et al.\ 1996; Folha \& Emerson 2001).  In the context of 
T Tauri stars, blue-shifted Br$\gamma$ centroids are 
taken to indicate that the Br$\gamma$ emission arises in 
gas infalling in a stellar magnetosphere 
(e.g., Muzerolle et al.\ 1998).
The line flux and emission centroid measured for V1331 Cyg  
are similar to the values reported earlier for the same object. 
We earlier found a line flux of $2.3\times 10^{-16}{\rm \,W\,m^{-2}}$
and an emission centroid blue-shifted by $-12\kms$
(Najita et al.\ 1996).

\clearpage
  
\begin{figure}
\epsscale{.9}
\plotone{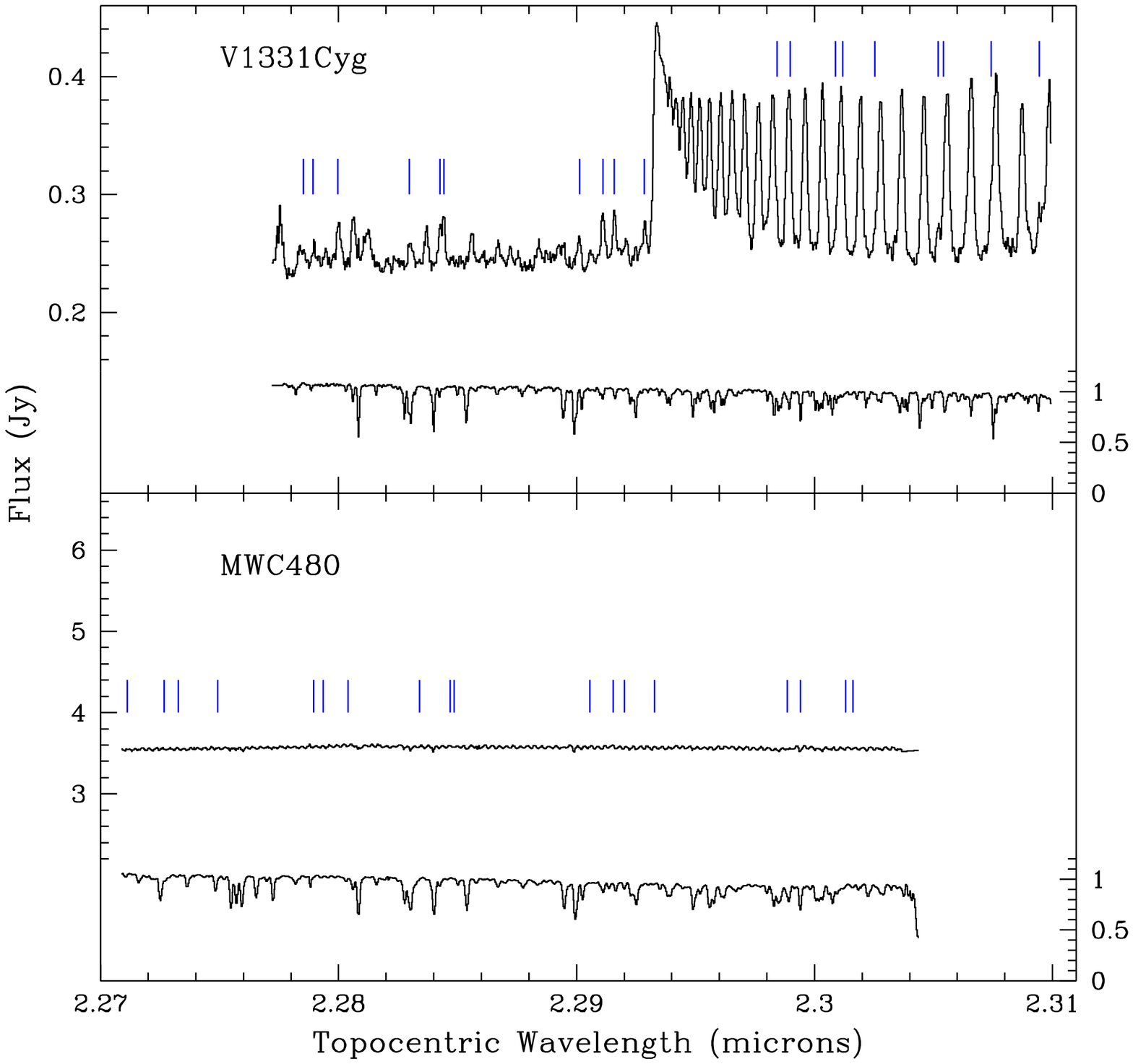}
\caption{Spectra of V1331 Cyg and MWC 480 in the $K$-band 
orders 33-38 are shown in Fig.~1a-1f, respectively. 
In each figure, the top panel shows the spectrum of V1331 Cyg 
(upper histogram) and the corresponding telluric standard 
spectrum (lower histogram).  The telluric spectrum 
is normalized to unity and plotted against the vertical scale on 
the right side of the plot.
The bottom panel of each figure shows the spectrum of MWC 480 
(upper histogram) and its corresponding telluric standard 
spectrum (lower histogram).  
In both panels, the positions of known water lines (Polyansky et al.\ 1997), 
shifted to the known stellar velocity in the topocentric frame 
of each object, are also indicated (vertical ticks); 
this required a velocity shift of $-28.7\kms$ and $+27.7\kms$ 
for the lines shown in V1331 Cyg and MWC 480, respectively.  
In Fig.~1c, the bracketted horizontal lines indicate the regions 
in which telluric features were not removed;  
the vertical dashed lines indicate the 
rest velocity of the Br$\gamma$ line at the expected stellar radial 
velocity of the source in the observed frame. 
False emission features near $1.979\micron$ and $2.144\micron$ 
in the MWC 480 spectrum (indicated by asterisks) are caused by 
uncorrected stellar absorption in the A7 telluric standard.
}
\end{figure}
\clearpage
{\plotone{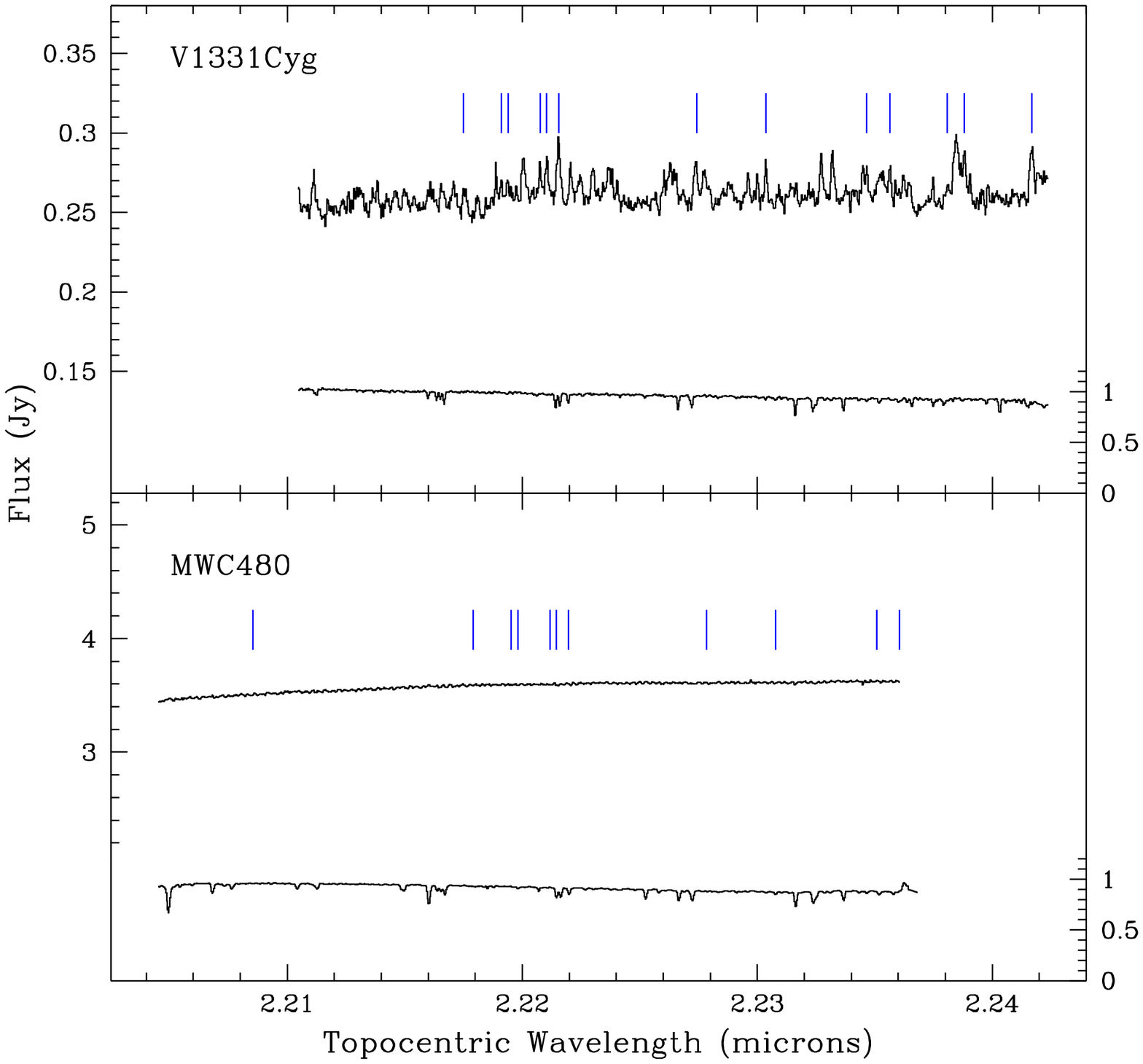}}\\
\centerline{Fig. 1b. --- Continued.}
\clearpage
{\plotone{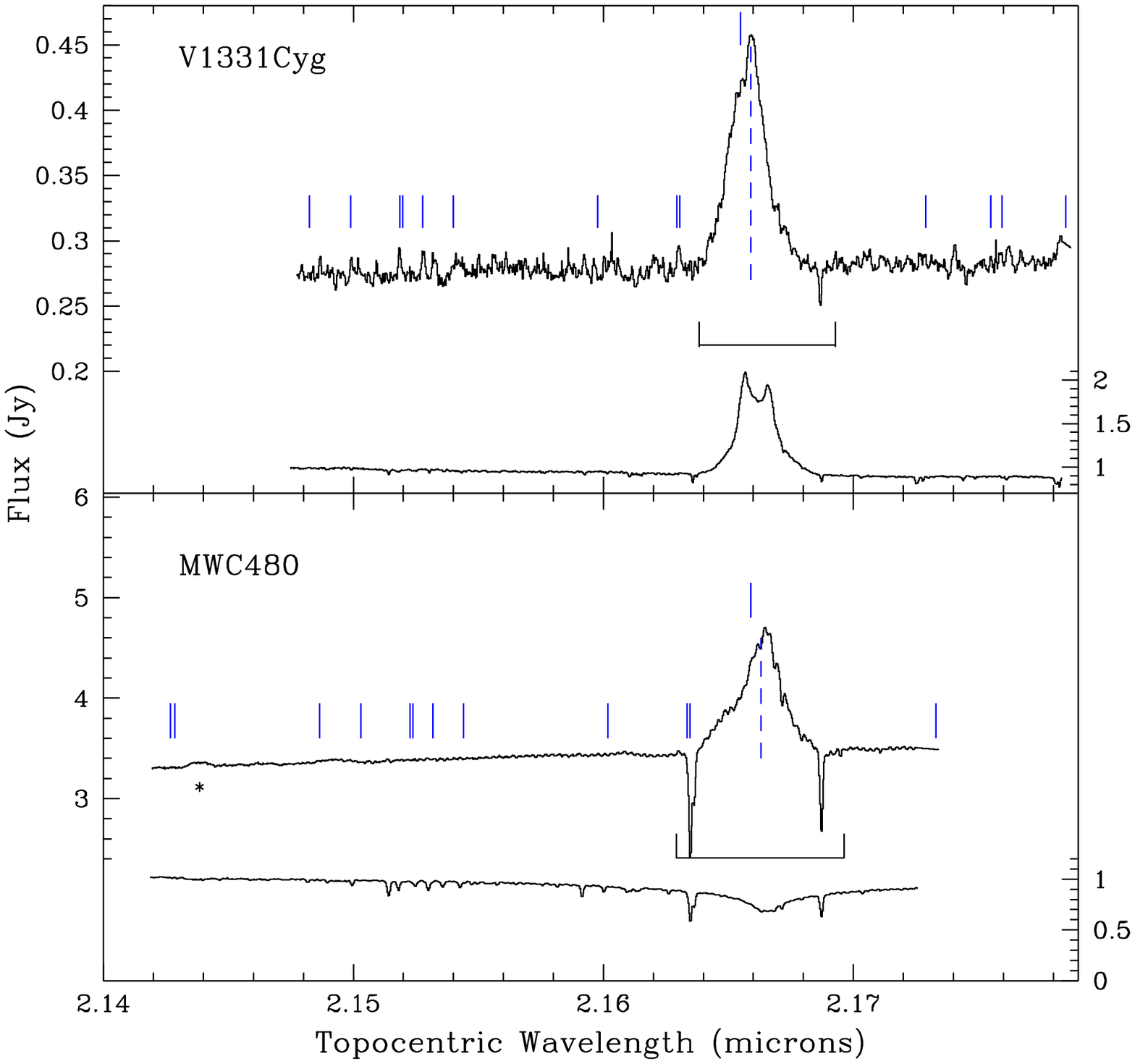}}\\
\centerline{Fig. 1c. --- Continued.}
\clearpage
{\plotone{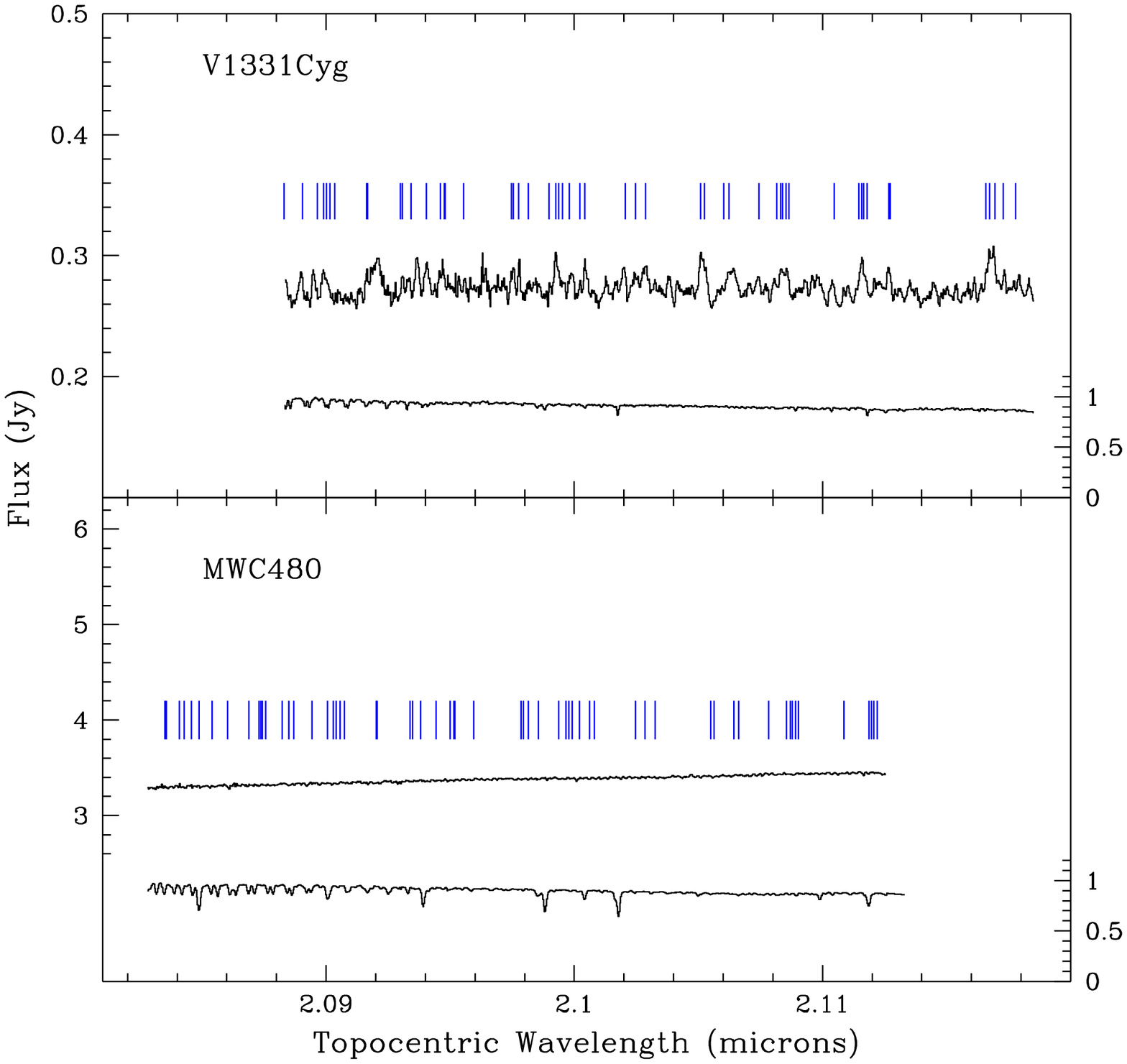}}\\
\centerline{Fig. 1d. --- Continued.}
\clearpage
{\plotone{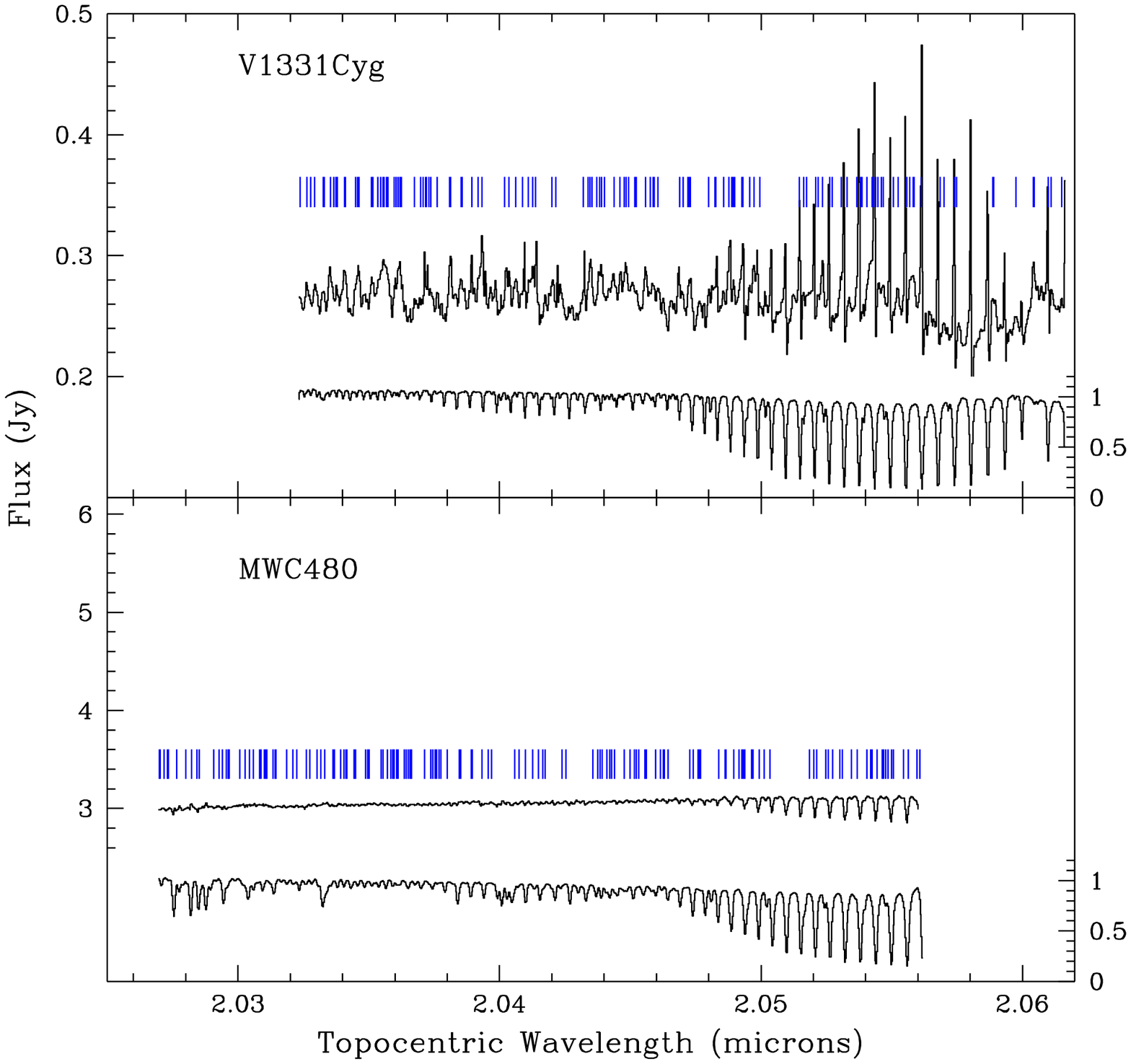}}\\
\centerline{Fig. 1e. --- Continued.}
\clearpage
{\plotone{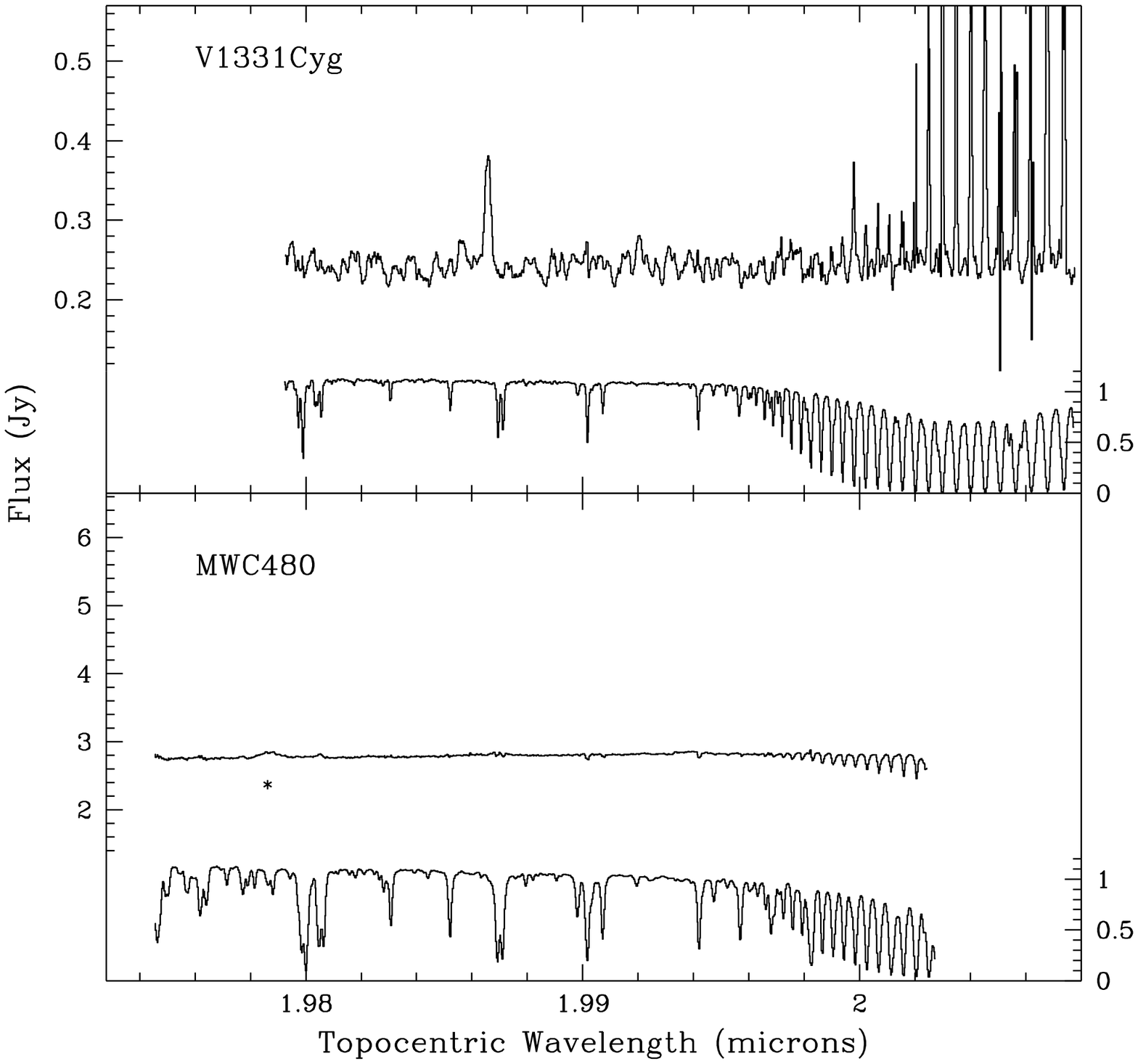}}\\
\centerline{Fig. 1f. --- Continued.}
\clearpage

\begin{figure}
\epsscale{0.9}
\plotfiddle{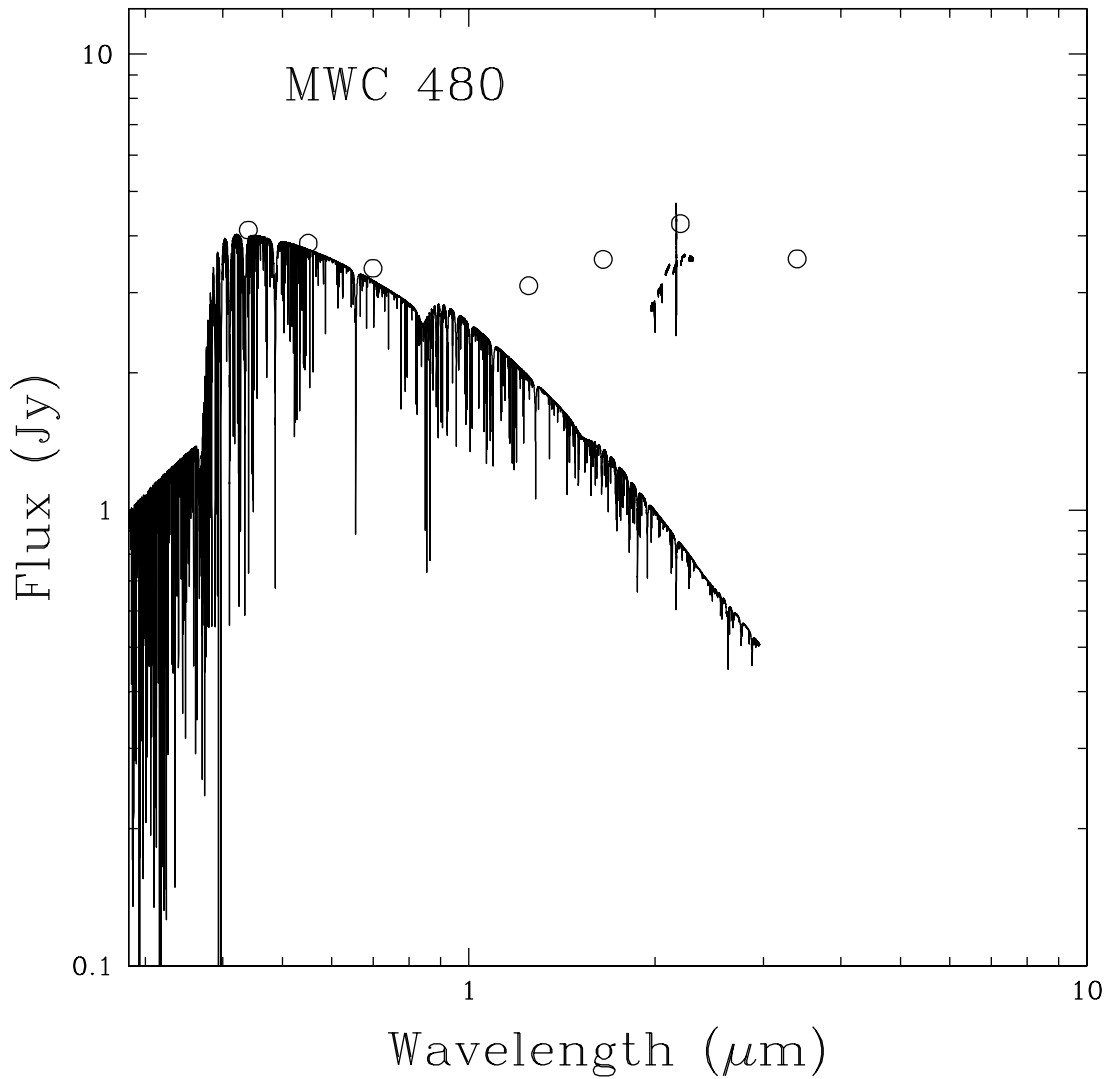}{5.5truein}{0}{75}{75}{-230}{-50}
\caption{Dereddened spectral energy distribution of MWC 480 (circles) 
compared 
with a Kurucz model for $T_{\rm eff}= 8400$\,K, $\log g =4.5$, 
a stellar luminosity of $22\Lsun$, 
and a distance of 170\,pc, 
assuming an extinction of $A_V=0.3$. 
Our flux calibrated NIRSPEC spectrum is also shown. 
See text for details.
}
\end{figure}

\begin{figure}
\epsscale{0.9}
\plotfiddle{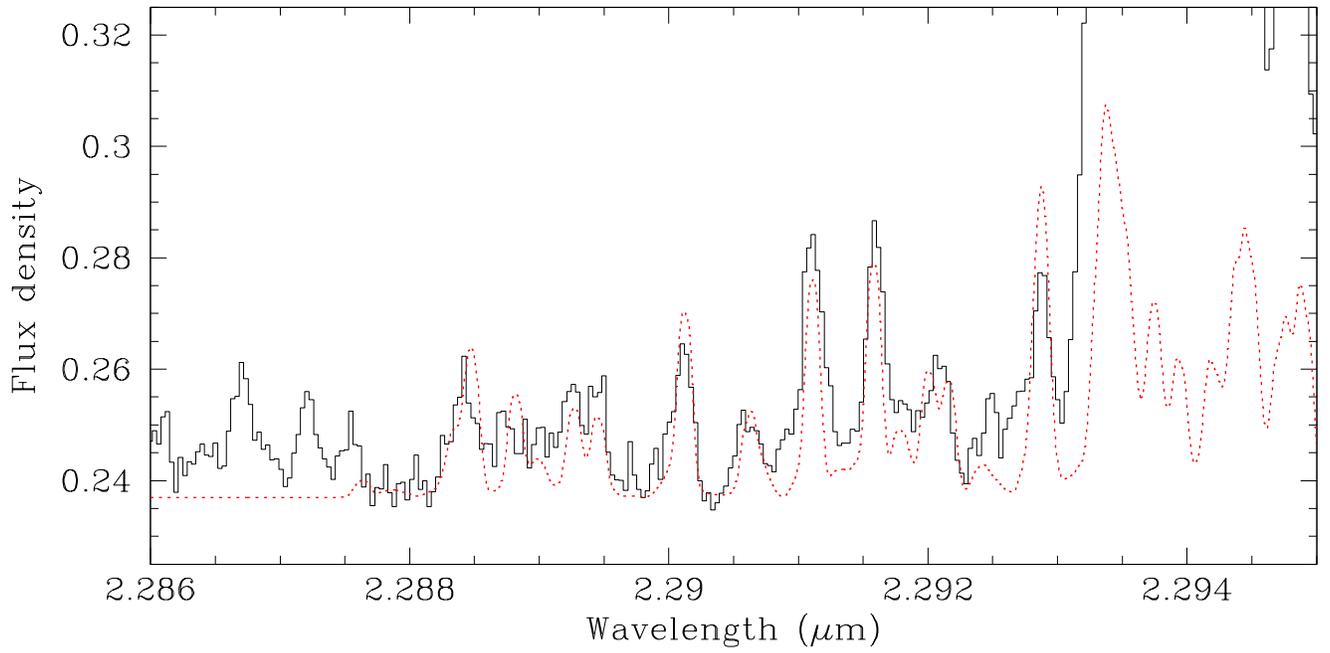}{5.5truein}{270}{75}{75}{-290}{430}
\caption{Observed spectrum of V1331 Cyg in the region blueward 
of the $v$=2--0 CO bandhead (black histogram) along with a  
synthetic disk spectrum that provides a fit to the water emission 
from this region (dotted line).  
(The water line list covers a limited spectral region, 
from  $\sim 2.287\micron$ to $\sim 2.30\micron$.) 
}
\end{figure}

\begin{figure}
\epsscale{0.9}
\plotfiddle{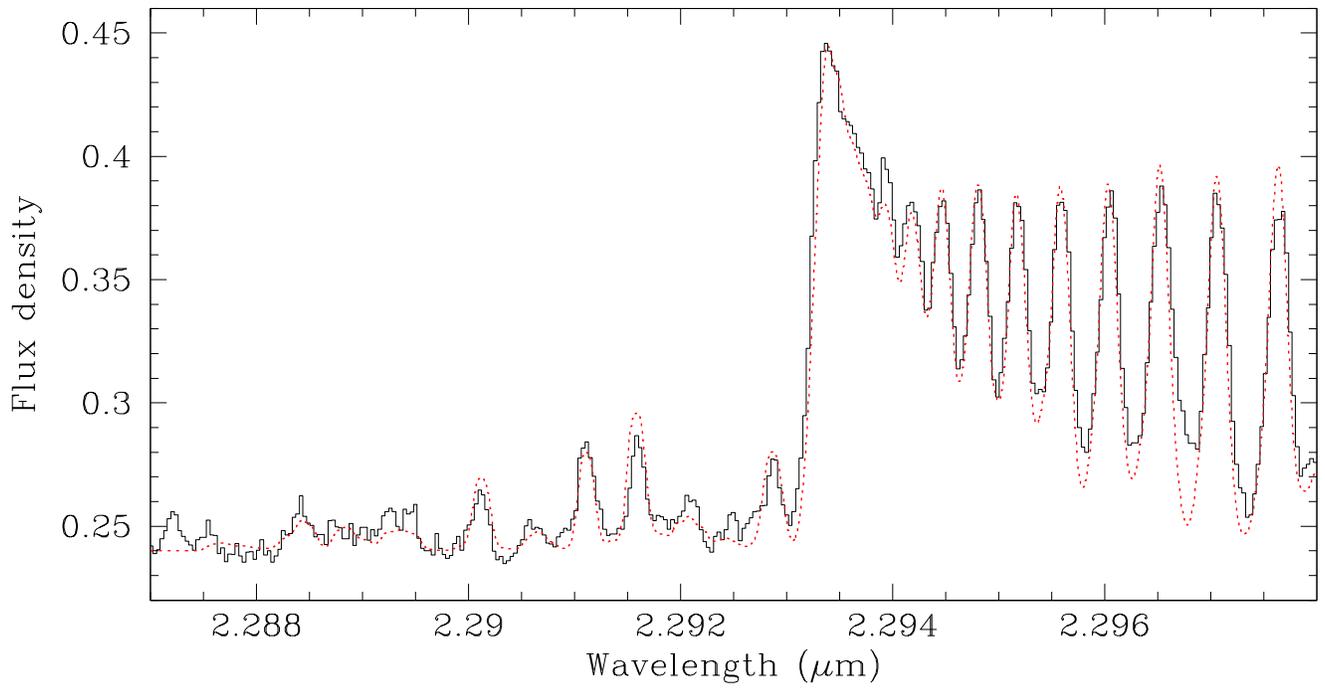}{5.5truein}{270}{75}{75}{-290}{430}
\caption{Observed spectrum of V1331 Cyg in the $2.3\micron$ region  
(black histogram).  A model in which the CO-to-water abundance 
ratio is much lower than in chemical equilibrium (dotted line)
provides a fit to the water and CO emission from this region 
(see text for details). 
}
\end{figure}

\begin{figure}
\epsscale{0.9}
\plotfiddle{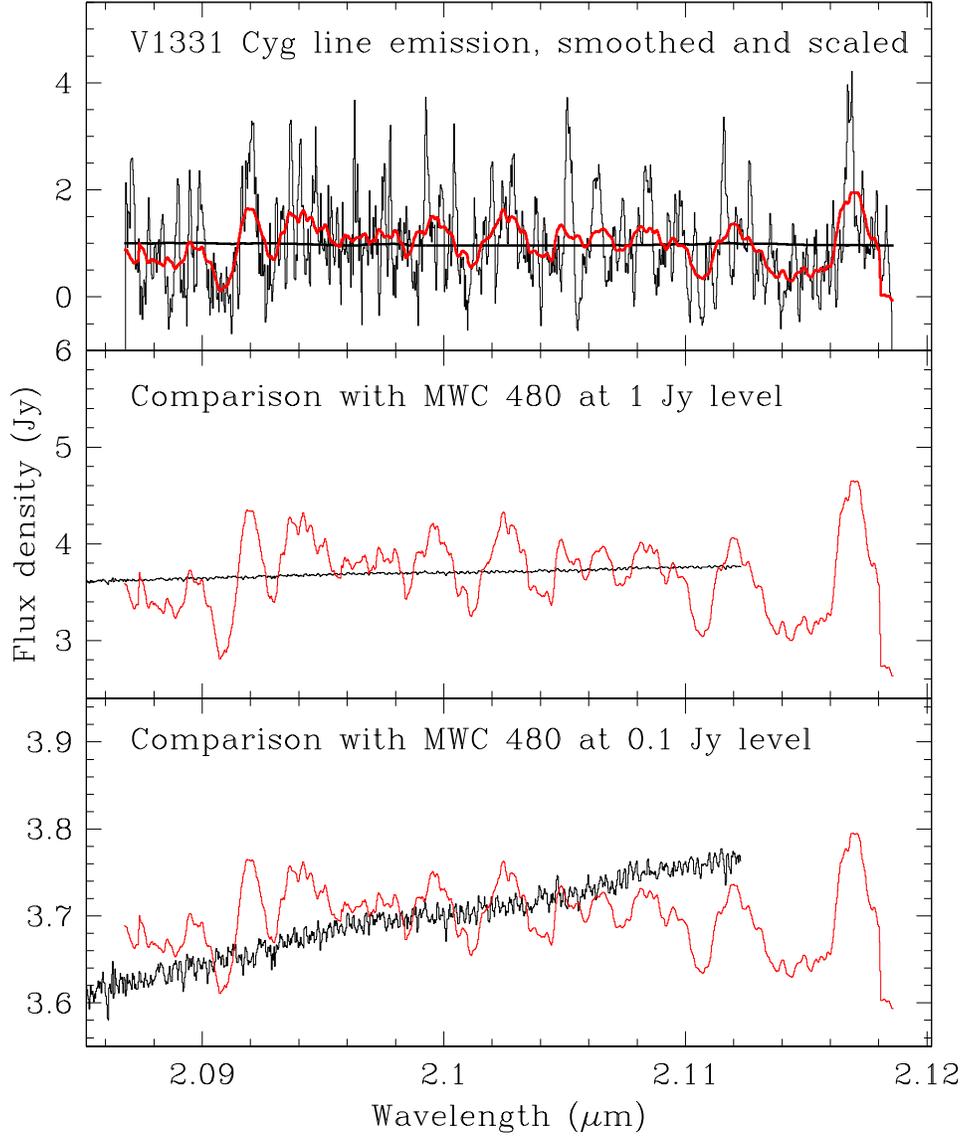}{5.7truein}{0}{70}{70}{-215}{-70}
\caption{{\it Top:} The continuum-subtracted line emission spectrum 
from V1331 Cyg (black histogram), 
boxcar smoothed by 
36 pixels (or $160\kms$; red line), 
twice the width of the CO fundamental emission from MWC 480, 
and normalized to an average intensity of unity across the 
order (horizontal line). 
{\it Middle:}  
The observed spectrum of MWC 480 (black line), compared with 
the smoothed spectrum of V1331 Cyg from the top panel (red line) 
which is normalized to the 1\,Jy strength of the hot compact 
continuum in the Eisner (2007) model of MWC 480. 
{\it Bottom:} 
As in the middle panel, but with the smoothed spectrum of 
V1331 Cyg (red line) normalized to the 0.1\,Jy strength of the 
water emission in the Eisner (2007) model of MWC 480. 
Spectral structure similar to that in the smoothed V1331 Cyg 
spectrum is not present in the MWC 480 spectrum at either 
the 1\,Jy or 0.1\,Jy level.
}
\end{figure}

\end{document}